%% file: paper.tex
\documentstyle[preprint,times,psfig]{infocom}
\include{html}

\begin{document}
%
% title page and abstract
%
\title{A Control and Management Network for Wireless ATM Systems}
\affiliation{Information and Telecommunication Technologies Center \\
        Department of Electrical Engineering \& Computer Science\\
        University of Kansas\\
        Lawrence, KS 66045-2228}
\support{The author can be contacted via e-mail at sbush@ittc.ukans.edu.
This work is partially funded by DARPA under contract J-FBI-94-223 and
Sprint under contract CK5007715.}
\author{\htmladdnormallink{Stephen F. Bush}
{http://www.ittc.ukans.edu/\~{}sbush}, Sunil Jagannath, Ricardo Sanchez, 
Joseph B. Evans, Victor S. Frost, Gary J. Minden and K. Sam Shanmugan
}
%\from
%Information and Telecommunications Technologies Center \\
%Department of Electrical Engineering \& Computer Science \\
%University of Kansas, Lawrence, KS 66045-2228 \\
%\email{sbush@ittc.ukans.edu}                                       
%}
%
\copyrightnotice{ACM/Baltzer Science Publishers}
%
%\toappear{\it{Journal of Wireless Networks} August 1997}
%
%\markright{S. Bush et Al.
%/ A Control and Management Network for Wireless ATM Systems}
%
%
\maketitle
\input{abstract}
%
% main body of paper
%

\input{s1}	% Introduction

% System Description

\input{s2}	% High Speed Network Overview
\input{s14}	% Physical Layer
\input{s13}	% Link Layer
\input{s12}	% ATM Layer
\input{s15}	% Network Layer (IP)
\input{s3}	% Network Control Protocol (NCP) Packets
\input{s4}	% NCP
\input{s11}	% ATM Network Configuration Layer
\input{s5}	% VNC
\input{w5}	% VNC+PNNI

% System Implementation & test results

\input{s9} 	% Development and Implementation
\input{s7}	% Packet Timing Results
\input{s10}	% Orderwire Bandwidth

% System Emulation results

\input{w2}	% NCP Analysis
\input{w3}	% Emulation Design
\input{w4}	% Emulation Results
\input{s6}	% Summary

%
% bibliography
%
\medskip
\bibliographystyle{unsrt} 
\bibliography{/home/bushsf/ref/standards,/home/bushsf/ref/topo,/home/bushsf/ref/psim,/home/bushsf/ref/mob,/home/bushsf/ref/twe}
\end{document}

%% file: html.tex
\ifx \htmlstyloaded\relax  \else\let\htmlstyloaded\relax\fi

\newcommand{\htmladdnormallink}[2]{#1}

\newcommand{\htmladdimg}[1]{}

\newcommand{\externallabels}[2]{}

\newcommand{\externalref}[1]{}

\makeatletter
\def\makeinnocent#1{\catcode`#1=12 }
\def\csarg#1#2{\expandafter#1\csname#2\endcsname}

\def\ThrowAwayComment#1{\begingroup
    \def\CurrentComment{#1}%
    \let\do\makeinnocent \dospecials
    \makeinnocent\^^L% and whatever other special cases
    \endlinechar`\^^M \catcode`\^^M=12 \xComment}
{\catcode`\^^M=12 \endlinechar=-1 %
 \gdef\xComment#1^^M{\def\test{#1}
      \csarg\ifx{PlainEnd\CurrentComment Test}\test
          \let\html@next\endgroup
      \else \csarg\ifx{LaLaEnd\CurrentComment Test}\test
            \edef\html@next{\endgroup\noexpand\end{\CurrentComment}}
      \else \let\html@next\xComment
      \fi \fi \html@next}
}
\makeatother

\def\includecomment
 #1{\expandafter\def\csname#1\endcsname{}%
    \expandafter\def\csname end#1\endcsname{}}
\def\excludecomment
 #1{\expandafter\def\csname#1\endcsname{\ThrowAwayComment{#1}}%
    {\escapechar=-1\relax
     \csarg\xdef{PlainEnd#1Test}{\string\\end#1}%
     \csarg\xdef{LaLaEnd#1Test}{\string\\end\string\{#1\string\}}%
    }}

\excludecomment{comment}
\excludecomment{rawhtml}

\excludecomment{htmlonly}
\newcommand{\html}[1]{}

\newcommand{\hyperref}[4]{#2\ref{#4}#3}

\newcommand{\htmlimage}[1]{}

\newcommand{\htmladdtonavigation}[1]{}

%% file: abstract.tex
\begin{abstract}

This paper describes the design of a control and management network 
(orderwire) for a mobile wireless Asynchronous Transfer Mode (ATM) 
network. This mobile wireless ATM network is part of the Rapidly 
Deployable Radio Network (RDRN).
The orderwire system consists of a packet radio
network which overlays the mobile wireless ATM network, each
network element in this network uses Global Positioning System (GPS) 
information to control a beamforming antenna
subsystem which provides for spatial reuse.
This paper also proposes a novel Virtual Network Configuration (VNC) 
algorithm for predictive network configuration. A mobile ATM Private 
Network-Network Interface (PNNI) based on VNC is also discussed. Finally, 
as a prelude to the system implementation, results 
of a Maisie simulation of the orderwire system are discussed.

\end{abstract}

%% file: s1.tex
\section{Introduction}

Research involving mobile wireless ATM is advancing rapidly.
One of the earliest proposals for a wireless ATM
architecture is described in \cite{Raychaudhuri}. 
In this paper, various alternatives for a wireless Media Access Channel
(MAC) are discussed and a MAC frame is proposed. The MAC
contains sequence numbers, service type, and a Time of Expiry (TOE) 
scheduling policy as a means for improving real-time data traffic
handling. A related work which considers changes to Q.2931
\cite{Q2931} to support mobility is proposed in \cite{Yuan}.
A MAC protocol for wireless ATM is examined in \cite{McTiffin}
with a focus on Code Division Multiple Access (CDMA) in which 
ATM cells are not preserved allowing a more efficient form of 
packetization over the wireless network links. The ATM cells
are reconstructed from the wireless packetization method after
being received by the destination. The Rapidly Deployable Radio Network 
Project (RDRN) architecture described in this paper maintains standard 
ATM cells through the wireless links. 
Research work on wireless ATM LANs have been described in \cite{Rednet} 
and \cite{SWAN}. The  mobile wireless ATM RDRN differs from these
LANs because the RDRN uses point-to-point radio communication over much 
longer distances.
The system described in \cite{BAHAMA} and \cite{Veer} consists of Portable 
Base Stations (PBS) and mobile users. PBSs are base stations which perform 
ATM cell switching and are connected via Virtual Path Trees which are 
preconfigured ATM Virtual Paths (VP). These trees can change based on 
the topology as described in the {\em Virtual Trees Routing Protocol} 
\cite{VTRP}. However, ATM cells are forwarded along the Virtual Path
Tree rather than switched, which differs from the ATM standard. An 
alternative mobile 
wireless ATM system is presented
in this paper which consists of a mobile PNNI architecture based on a 
general purpose predictive mechanism known as Virtual Network Configuration 
that allows seamless rapid handoff.

The objective of the Rapidly Deployable Radio Network (RDRN) effort 
is to create an ATM-based wireless communication system that 
will be adaptive at both the link and network levels to allow for rapid 
deployment and response to a changing environment. The objective of the 
architecture is to use adaptive point-to-point topology to gain the 
advantages of ATM for wireless networks. A prototype of this system has 
been implemented and will be demonstrated over a wide area
network. The system adapts to its environment and can automatically arrange 
itself into a high capacity, fault tolerant, and reliable network. 
The RDRN architecture is composed of two overlaid networks: 

\begin{itemize}
\item a low bandwidth, low power omni-directional network for location 
dissemination, switch coordination, and management which is the orderwire 
network described in this paper,
\item a ``cellular-like'' system for multiple end-user access to the switch
using directional antennas for spatial reuse, and 
and a high capacity, highly directional, multiple beam network for 
switch-to-switch communication.
\end{itemize}

\begin{figure*}[htbp]
        \centerline{\psfig{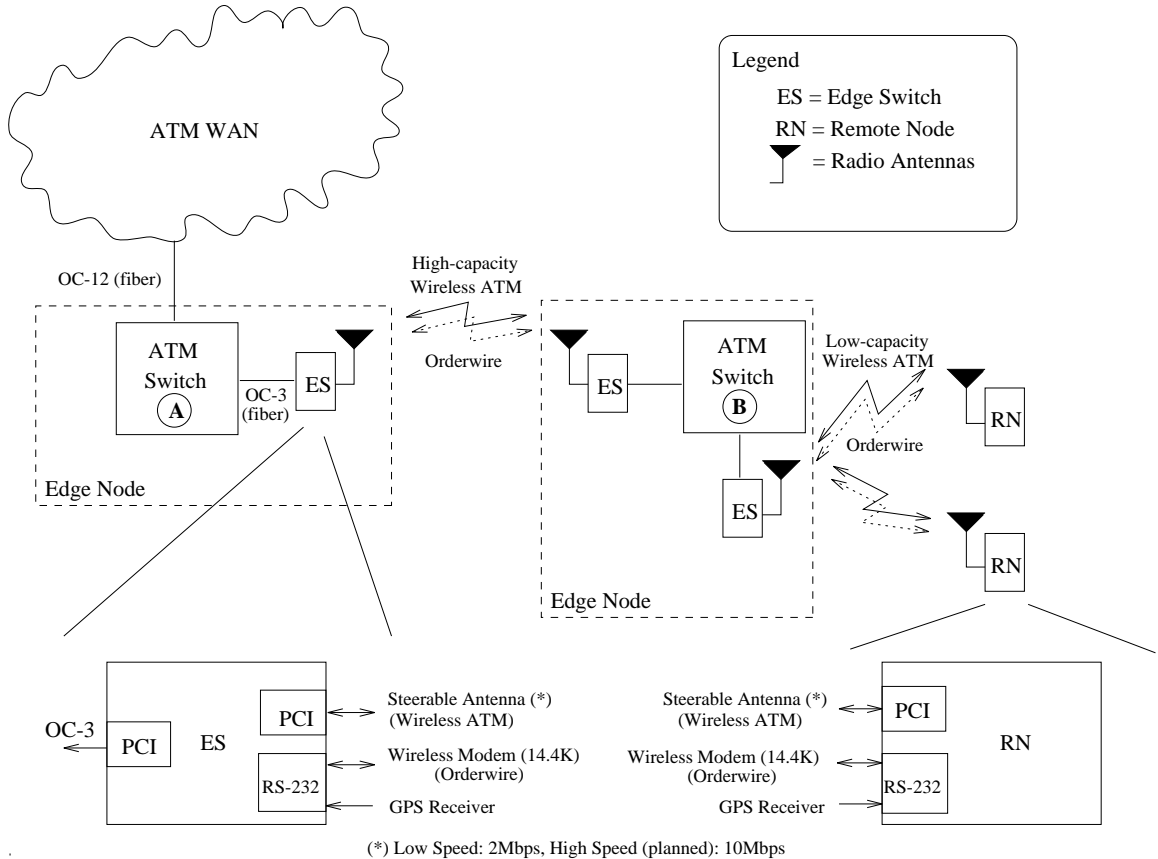}}
        \caption{RDRN High-level Architecture.}
        \label{scenario}
\end{figure*}

\begin{figure*}[htbp]
        \centerline{\psfig{file=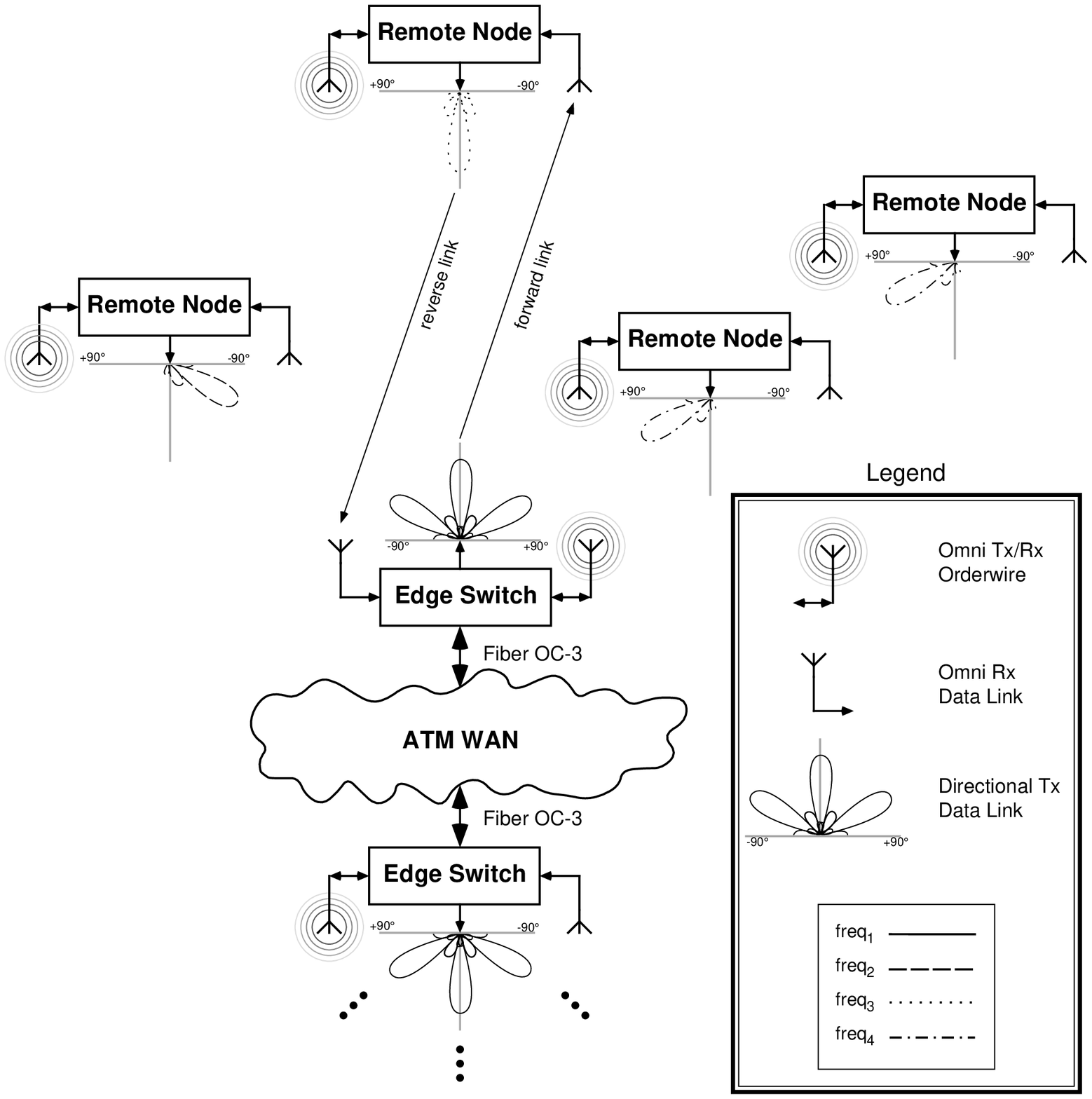,width=6.0in}}
        \caption{RDRN Component Overview.}
        \label{hwowov}
\end{figure*}

The network currently consists of two types of nodes, Edge Nodes (EN) and
Remote Nodes (RN) as shown in Figure \ref{scenario}. ENs where designed to
reside on the edge of a wired network and provide access to the wireless 
network; however, EN also has wireless links.
The EN components include Edge Switches (ES) and optionally an ATM switch, 
radio handling the ATM-based communications, packet radio for the low speed 
orderwire running a protocol based on X.25 (AX.25), GPS receiver, and a 
processor. Host nodes or remote nodes (RN) consist of the above, but do 
not contain an ATM switch.
The ENs and RNs also include a phased array steerable antenna. The RDRN uses
position information from the GPS for steering antenna beams toward nearby 
nodes and nulls toward interferers, thus establishing the high capacity links
as illustrated in Figure \ref{hwowov}. Figure \ref{hwowov} highlights an
ES (center of figure) with its omni-directional transmit and receive 
orderwire antenna and an omni-directional receive and directional transmit 
ATM-based links. Note that two RNs share the same $45^o$ beam from the
ES and that four distinct frequencies are in use to avoid interference.
The decision involving which beams to establish and which frequencies
to use is made by the topology algorithm which is discussed in a later
section.

The ES has the capability of switching ATM cells among connected RNs or 
passing the cells on to an ATM switch to wire-based nodes. Note that the 
differences between an ES and RN are that the ES performs switching and 
has the capability of higher speed radio links with other Edge Switches 
as well as connections to wired ATM networks.

\begin{figure}[htbp]
        \centerline{\psfig{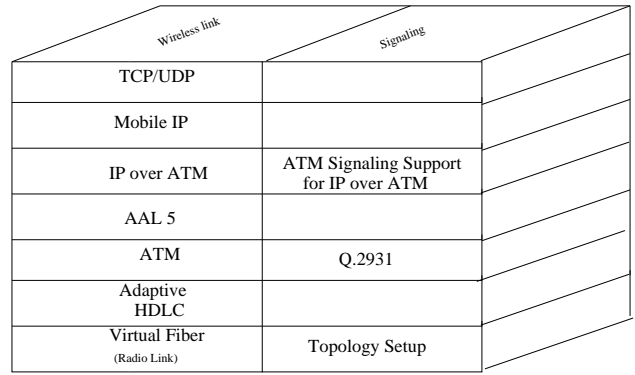}}
        \caption{Wireless ATM Protocol Stack.}
        \label{hsstack}
\end{figure}

The orderwire network uses a low power, omni-directional channel, operating at 
19200 bps, for signaling and communicating node locations to other network 
elements. The orderwire aids link establishment between the ESs and between 
the RNs and ESs, tracking  remote nodes and determining link quality. 
The orderwire operates over packet radios and is part of the Network
Control Protocol (NCP)\footnote{The Simple Network Management Protocol
(SNMP) Management Information Base (MIB) for the NCP operation as well as
live data from the running prototype RDRN system can be
retrieved from \htmladdnormallink{http:/\-/\-www.ittc.ukans.edu/\-$\sim$sbush/\-rdrn/\-ncp.html}{http://www.tisl.ukans.edu/~sbush/rdrn/ncp.html}.}.
An example of the user data and orderwire network topology is shown in
Figure \ref{owov}. In this figure, an ES serves as a link between
a wired and wireless network, while the remaining ESs act as wireless
switches. The protocol stack for this network is shown in 
Figure \ref{hsstack}.

The focus of this paper is on the NCP and in particular on the orderwire 
network and protocols. This includes protocol layer 
configuration, link quality, hand-off, and host/switch assignment
along with information provided by the GPS system such as position
and time. The details of the user data network will be covered in 
this paper only in terms of services required from, and interactions 
with, the NCP. 

Section \ref{WatmReqS} provides a more detailed description of the RDRN 
system, with a focus on the requirements and interaction of each protocol 
layer with the NCP. Operation of the NCP is described in Section \ref{initimp}. 
A new concept known as Virtual Network Configuration (VNC) is explained 
in Section \ref{vnc} along with an example application of
a Mobile Private Network-Network Interface (PNNI) enhanced with VNC. The 
development and implementation of the NCP is described along with initial 
timing results in Section \ref{emulation}. In Section \ref{NCPanal},
an analysis of NCP indicates the performance of NCP as the system is 
scaled up. Finally, emulation results are presented in Section \ref{emres}.

\begin{figure*}[htbp]
        \centerline{\psfig{file=figures/owov.eps,width=6.0in}}
        \caption{Example Orderwire Topology.}
        \label{owov}
\end{figure*}

%% file: s2.tex
\section{Wireless ATM-Based Network Configuration Requirements}
\label{WatmReqS}

This section provides a brief overview of the high speed 
protocol architecture for the RDRN wireless ATM network 
\cite{BushRDRN}.
The purpose is to introduce the RDRN network
and more importantly to identify the requirements that each layer
will have for the network configuration protocol.

%% file: s14.tex
\subsection{Physical Layer}
\label{phys}

The physical layer includes all hardware components and the wireless
connections. This includes the high speed radios, orderwire packet
radios, ATM switch, antennas, and additional processor for
configuration and setup. This layer provides a raw pipe
for the data link layer described in the next section. 
Directional beams from a single antenna are used to
obtain spatial reuse and Time Division Multiple Access (TDMA) is used 
to provide access to multiple RNs within a beam.
The physical layer details can be found in \cite{EwyRDRN}.
The NCP sets up the physical layer wireless connections. 

%% file: s13.tex
\subsection{Link Layer}
\label{link}

In this architecture ATM will be carried end-to-end. However, at the
edge between the wired (high-speed) network and wireless links,      
multiple ATM cells will be combined into an HDLC-like frame. These
frames comprise the Adaptive HDLC (AHDLC) protocol.
The wireless data link layer
is adaptive to provide an appropriate trade-off between data rate and
reliability in order to support the various services.
For example, we may want to drop voice packets, which are 
time sensitive, but retry data packets. The edge interface unit makes
this decision based on 
knowledge of the requirements of each traffic stream, possibly based
on virtual circuit number.

For some types
of traffic, error correction may be achieved using
retransmission. Here, delay is increased for this class of traffic to
prevent cell losses. It is well known that even a few cell losses can
have a significant impact on the performance of TCP/IP 
(Transmission Control Protocol/Internet Protocol), while TCP/IP
can cope with variable delays \cite{Caceres}. The 
Adaptive High Level Data Link Control (AHDLC) protocol can 
change in response to traffic requirements. ATM end-to-end provides 
the following benefits:
\begin{enumerate}
\item Moderate cut-through, e.g. an IP segment may contain 8192
bytes or about 170 cells, while one ATM HDLC-like frame will contain on the
order of 3-20 cells
\item ATM is a standard protocol.
\item ATM can incorporate standardized Quality of Service (QoS) parameters which 
could be based on the virtual circuit identifier.
\end{enumerate}

The link layer must also maintain cell order; this will be critical
during hand-off of an RN from one ES to another. Details of the
Adaptive HDLC protocol and frame structures can be found in
\cite{BushRDRN} and \cite{EwyRDRN} .

%% file: s12.tex
\subsection{ATM Layer}
\label{ATM}

The protocol on the Edge Switch (ES) will remove
ATM cells from the AHDLC frames and switch them to 
the proper port. It will also
pack ATM cells into an Adaptive HDLC frame to send to the radio. 
The ATM Device Driver API and Adaptive Driver are detailed in
\cite{BushRDRN}. Note that standard ATM call setup signaling is used 
and no AAL is precluded from use.

%% file: s15.tex
\subsection{Network Layer}
\label{ip}

This section of the architecture is concerned with the Internet
Protocol and how it relates to ATM and mobility. This layer provides a
well known and widely used network layer, whose primary purpose is to
provide routing between subnetworks and service for the 
Transmission Control Protocol (TCP) and User Datagram Protocol (UDP)
transport layers. The relation between IP and ATM is still an open
issue. {\it Classical IP and ARP over ATM} \cite{RFC1577} (CLIP) is an 
initial standard solution. However, it has several weak points such as
requiring a router to connect Logical IP Subnetworks (LIS) even when
they are directly connected at the ATM level, and requiring an ATM 
Address Resolution Protocol (ARP) server to provide address resolution 
for a single LIS. The
{\it Non-Broadcast Multiple Access (NBMA) Next Hop Resolution Protocol}
(NHRP)\cite{NBMA} provides a better solution but it is still in
draft form.
The RDRN architecture has implemented CLIP and supports both PVCs and
SVCs via ATMARP.

%% file: s3.tex
\section{Network Control Protocol Overview}
\label{initimp}

An initial implementation of the RDRN Network Control Protocol (NCP) 
for the prototype system is presented next. The physical layer of the high
speed radio connection has a corresponding layer in the NCP, as 
shown in Table \ref{prototab}. The following is a description and ordering 
of events for the establishment of the wireless connections.

\begin{table*}[htbp]
\centering
\begin{tabular}{||l|l|l||}	                                   \hline
{\bf Protocol Layer} & {\bf Packet Types}  & {\bf Packet Contents}         \\ \hline \hline
Physical Layer	& MYCALL  	& Callsign, Start-Up-Time                  	\\ \hline
	& NEWSWITCH	& {\em empty packet} 			   		\\ \hline
	& SWITCHPOS   	& GPS Time, GPS Position 			\\ \hline
	& TOPOLOGY 	& Callsigns and Positions of each node	 	\\ \hline
	& USER\_POS	& Callsign, GPS Time, GPS Position             	\\ \hline
	& HANDOFF         & Frequency, Time Slot, ES GPS Position 				\\ \hline
\end{tabular}
\caption{\label{prototab} Network Control Protocol Packets.}
\end{table*}

%% file: s4.tex
\subsection{Physical Layer of the Network Control Protocol}
\label{owphys}
  
At the physical level we will be using the orderwire to exchange
position, time and link quality information and to setup the wireless
connections. The process of setting up the wireless connections
involves setting up links between ESs and between ESs and RNs.

The network will have one master ES, which will run
the topology configuration algorithm \cite{cla} and distribute the resulting
topology information to all the connected ESs over point-to-point
orderwire packet radio links. In the current prototype the point-to-point link layer 
for the orderwire uses AX.25 \cite{AX.25}. 
The master ES is initially the first active ES,
and any ES has the capability of playing the role of the
master.

The first ES to become active initially broadcasts its
callsign and start-up-time in a {\bf MYCALL} packet, and listens for responses 
from any other ESs. In this prototype system, the packet radio callsign is 
assigned by the FCC and identifies the radio operator.
Since it is the first active ES, there would be no responses in a given time 
period, say T. At the end of T seconds, the ES rebroadcasts its {\bf MYCALL}
packet and waits another T seconds. At the end of
2T seconds, if there are still no responses from other ESs, the
ES assumes that it is the first ES active and takes on
the role of the master. If the first two or more ESs start up
within T seconds of each other, at the end of the interval T, the ESs
compare the start-up times in all the received
{\bf MYCALL} packets and the ES with the oldest start-up time
becomes the master. In this system, accurate time stamps are provided by 
the GPS.

\begin{figure}[htbp]
        \centerline{\psfig{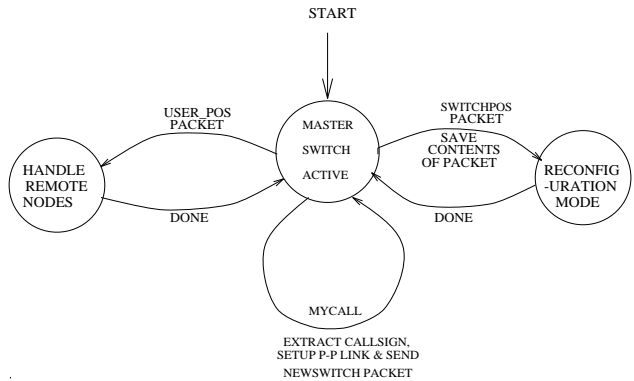}}
        \caption{State Diagram for Master EN.}
        \label{pap1}
\end{figure}

\begin{figure}[htbp]
        \centerline{\psfig{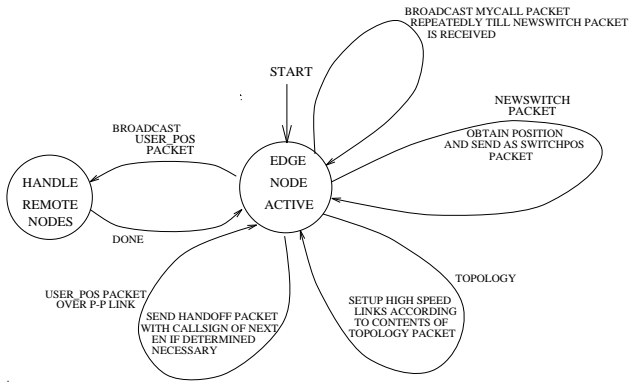}}
                \caption{State Diagram for EN not serving as Master.}
        \label{pap3}
\end{figure}

Each successive ES that becomes active initially
broadcasts its callsign in a {\bf MYCALL} packet. The master on
receipt of a {\bf MYCALL} packet extracts the callsign of the source,
 establishes a point-to-point link to the new ES and
sends it a {\bf NEWSWITCH} packet. The new ES on 
receipt of the {\bf NEWSWITCH} packet over a point-to-point orderwire link, obtains 
its position from its GPS receiver and sends its position to the master as a
{\bf SWITCHPOS} packet over the point-to-point orderwire link. 
On receipt of a {\bf SWITCHPOS} packet, the master records the position 
of the new ES in its switch position table, which is a table of ES positions, and 
runs the topology configuration algorithm \cite{cla} to determine the best 
possible interconnection of all the ESs. The master then distributes the 
resulting information to all the ESs in the form of a {\bf TOPOLOGY} packet 
over the point-to-point orderwire 
links. Each ES then uses this information to setup the
inter-ES links as specified by the topology algorithm. The master 
also distributes a copy of its switch position table to all the ESs
over the point-to-point orderwire links, which they can use in configuring RNs
 as discussed below. This sequence of operations is illustrated in Figure
 \ref{pap1} and Figure \ref{pap3}. 
Also, the ES then uses the callsign information in the
switch position table to setup any additional point-to-point orderwire
packet radio 
links corresponding to the inter ES links required to exchange any link
quality information. Thus this scheme results in 
a point-to-point star
network of orderwire links with the master at the center of the star and also point-to-point
orderwire links between
those ESs that have a corresponding inter ES link, as shown in Figure 
\ref{owov}.

In the event of failure of the master node which can be detected by listening
for the AX-25 messages generated on node failure, the remaining ESs exchange 
{\bf MYCALL} packets, elect a new master node, and the network of ESs is 
reconfigured using the topology configuration algorithm \cite{cla}.
 
%\begin{figure}[htbp]
%       \centerline{\psfig{file=figures/pap4.eps,width=3.25in}}
%        \caption{Flow Diagram for processes during Reconfiguration Mode State.}
%        \label{pap4}
%\end{figure}

%\begin{figure}[htbp]
%        \centerline{\psfig{file=figures/pap5.eps,width=3.25in}}
%        \caption{Flow Diagram for processes during Handle RN State.}
%        \label{pap5}
%\end{figure}

\begin{figure}[htbp]
        \centerline{\psfig{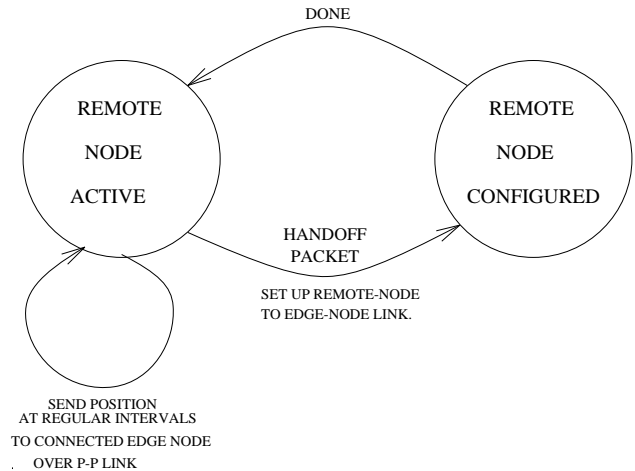}}
        \caption{State Diagram for RN.}
        \label{pap2}
\end{figure}

Each RN that becomes active obtains its position from
 its GPS receiver and broadcasts its position as a {\bf USER\_POS}
 packet over the orderwire network. This packet is received by all 
the nearby ESs. Each candidate ES then computes the distance between 
the RN and all the candidate ESs which is possible since each ES has the 
positions of all the other ESs from the switch position table. An initial
guess at the best ES to handle the RN is the closest ES. This ES
 then feeds the new RN's position information along with the
 positions of all its other connected RNs to a beamforming
 algorithm that returns the steering angles for each of the beams on the ES
 so that all the RNs can be configured.  If the beamforming algorithm 
determines that a beam and TDMA time slot
are available to support the new RN, 
the ES 
steers its beams so that all its connected RNs and the new
 RN are configured. It also records the new RN's position in its user position 
table which contains positions of connected RNs, establishes a 
point-to-point orderwire link to the new RN and sends it a {\bf HANDOFF} packet 
with link setup information indicating that the RN is 
connected to it. If the new RN cannot be accommodated, the ES sends it 
a {\bf HANDOFF} packet with the callsign of the next closest ES, to which 
the RN sends another {\bf USER\_POS} packet over a point-to-point
orderwire link. This ES then uses the beamform algorithm to
 determine if it can handle the RN. Figure \ref{pap2} shows the 
states of operation and transitions between the states for a RN.

This scheme uses feedback from the beamforming algorithm together with
 the distance information to configure the RN. It should be noted that the 
underlying AX.25 protocol \cite{AX.25} provides error free transmissions over 
point-to-point orderwire links. Also the point-to-point orderwire link can be 
established from either end and the handshake mechanism for setting up such 
a link is handled by AX.25. If the RN does
 not receive a {\bf HANDOFF} packet within a given time it uses a
 retry mechanism to ensure successful broadcast of its {\bf USER\_POS} 
packet.  

A point-to-point orderwire link is retained as long as a
 RN is connected to a particular ES and a
 corresponding high-speed link exists between them to enable exchange
 of link quality information. The link can be torn down when the
 mobile RN migrates to another ES in case of a
 hand-off. Thus at the end of this network configuration process, three overlaid
networks are setup, namely, an orderwire network, an RN to ES network and 
an inter-ES network. The orderwire network has links between the master ES and 
every other active ES in a star configuration, links between ESs connected by 
inter-ES links as well as links between RNs and the ESs to which they are connected, as
shown in Figure \ref{owov}.  
Raw pipes for the user data links between RNs and appropriate ESs as well as for the 
user data links between ESs are also set up. 

%% file: s11.tex
\subsection{ATM Network Configuration Layer}

This section briefly describes how ATM VCs are setup by the NCP.
As the orderwire network determines the topology
of all nodes in the wireless segment (e.g., RNs, ESs) in our architecture, 
and establishes 
link connectivity among adjacent nodes, setup is still required of the
actual ATM circuits on which wireless ATM are carried on the user
data overlay network. This is accomplished by providing standard
ATM signaling capabilities to RNs and ESs and using Classical IP over ATM
\cite{RFC1577} to associate ATM VCs to IP addresses. The Classical IP
over ATM implementation provided works for PVCs and SVCs (using ATMARP).
Since an ES may connect to multiple RNs (wireless connections) or 
ATM switches (wired connections), it can be thought of as a software-based 
ATM switch. 
In this sense, an ES features ATM PNNI signaling while an RN features 
ATM UNI signaling. By default, an RN creates one wireless-ATM protocol 
stack and establishes an ATM VC signaling channel on such a stack; however, 
the stack is initially in an inactive state (i.e., non-operational mode) 
since there is no link connectivity to another node established yet.
Likewise, an ES creates a predefined number of wireless-ATM protocol
stacks -- acting like ports in an ATM switch -- and establishes ATM VC
signaling channels on all configured stacks which are also initialized
as inactive. Wireless-ATM protocol stacks are controlled                  
by a daemon, called the adaptation manager, which acts on behalf of the 
orderwire network. The adaptation manager daemon not only controls the
stacks by setting their state to either active or inactive (default),
but also may modify configuration parameters of the stacks to
provide dynamic adaptation to link conditions.
Two possible scenarios illustrate the interactions
between the orderwire network and the wireless-ATM network.
In the first scenario the orderwire detects link connectivity between
an adjacent pair of nodes (e.g., RN-ES or ES-ES). In this case, 
the orderwire network requests an
inactive stack from the adaptation manager daemon at each end and
associates them with a designated address. Upon establishment of
link connectivity, a requested wireless stack has its state set to
active and is ready to operate. Note that since the signaling channels
are preconfigured on the stacks in question, users on the wireless
establish end-to-end connections exactly as if they were connected 
in a wired ATM network. The other scenario occurs when the orderwire
network detects a broken connection, at the link level, between  
two connected nodes. This case is typical of an RN moving away from the
connectivity range of an ES. The orderwire network thus contacts the
adaptation manager daemon at each end to set the wireless stacks
in question to inactive. Since a wireless stack is never destroyed, 
it can be reused in a future request from the orderwire to establish
connectivity to another pair of nodes.

%% file: s5.tex
\section{Virtual Network Configuration for a Rapidly Deployable Network}
\label{vnc}

In order to make RDRN truly rapidly deployable, configuration at all layers has
to be a dynamic and continuous process. Configuration can be a
function of such factors as load, distance, capacity and permissible 
topology, all of which are constantly changing in a mobile environment. 
A Time Warp \cite{Jefferson} based algorithm is used to anticipate 
configuration changes and speed the reconfiguration process.

\subsection{Virtual Network Configuration Algorithm}

The Virtual Network Configuration (VNC) algorithm is an application of a 
more general mechanism called Time Warp Emulation (TWE). Time Warp 
Emulation is a modification of Time Warp \cite{Jefferson}. 
The motivation behind TWE is to allow the actual components of a real-time 
system to work ahead in time in order to predict
future behavior and adjust themselves when that behavior does not
match reality. This is accomplished by realizing that there are now two
types of {\em false} messages, those which arrive in the past relative
to the process's Local Virtual Time (LVT) and those messages which have 
been generated which are time-stamped with the current real time, but 
whose values exceed some tolerance from the component's current value.

The basic Time Warp mechanism is modified by adding a verification 
query phase. This phase occurs when real time matches the receive time 
of a message in the output queue of a process. In this phase, the physical 
device being emulated in time is queried and the results compared with the value 
of the message. A value exceeding a prespecified tolerance will cause a
rollback of the process. 

\subsection{Virtual Network Configuration Overview}

The Virtual Network Configuration (VNC) algorithm can be explained
by an example.  A remote node's direction, velocity, bandwidth used, number
of connections, past history and other factors can be used to approximate a new
configuration sometime into the future. All actual configuration
processes can begin to work ahead in time to where the remote node is
expected to be at some point in the future. If the prediction is
incorrect, but not far off, only some processing will have to be
rolled back in time. For example, the beamsteering process results may have to
be adjusted, but the topology and many higher level requirements will still
be correct. Working ahead and rolling back to adjust for error
with reality is an on-going process, which depends on the tradeoff
between allowable risk and amount of processing time allowed into the
future. As a specific example, consider the effects of hand-off on TCP 
performance as described in \cite{Caceres}. In this work, throughputs
were measured for hand-off under various conditions and determined
to degrade badly.

\begin{comment}
\begin{table}[hbp]
\centering
\begin{tabular}{||l|l||}                                           \hline
{\bf Type of Hand-off}  & {\bf Bandwidth}         \\ \hline \hline
No Hand-offs            & 100\%                                  \\ \hline
Overlapping hand-offs   & 94\%                                  \\ \hline
0-second rendezvous delay & 88\%                                  \\ \hline
1-second rendezvous delay & 69\%                                  \\ \hline
\end{tabular}
\caption{\label{hoff} Bandwidth Loss Due to Hand-off.}
\end{table}
\end{comment}

\subsection{Virtual Network Configuration Implementation}

The effort required to enhance the network configuration algorithm
to include Virtual Network Configuration is minimal. Three new fields are 
added to each existing message in Table \ref{prototab}: antimessage toggle, 
send time, and receive time. Physical processes include beamforming, 
topology acquisition, table updates, and all processing required for 
configuration. Each physical process is assigned a tolerance. When
the value of a real message exceeds the tolerance of a predicted
message stored in the send queue, the process is rolled back.

Also, an additional packet type was created for updating an approximation 
of the Global Virtual Time (GVT). Because the system is composed of
asynchronously executing logical processes, each working ahead as
quickly as possible with its own local notion 
of time, it is necessary to calculate the time of the system as a whole.
This system-wide time is the GVT. The difference between GVT and 
current time is the amount of lookahead, $\Lambda$. Although $\mbox{GVT} 
\ge t$ where $t$ is real time, $\Lambda$ is required because 
it is used to control the efficiency and accuracy of the system. Since 
the network configuration 
system uses a master node as described in the physical layer setup, this 
is a natural centralized location for a centralized GVT update method. 
RNs transmit their LVT to the master, the master calculates an approximate 
GVT and returns the result.

\begin{figure*}[htbp]
        \centerline{\psfig{file=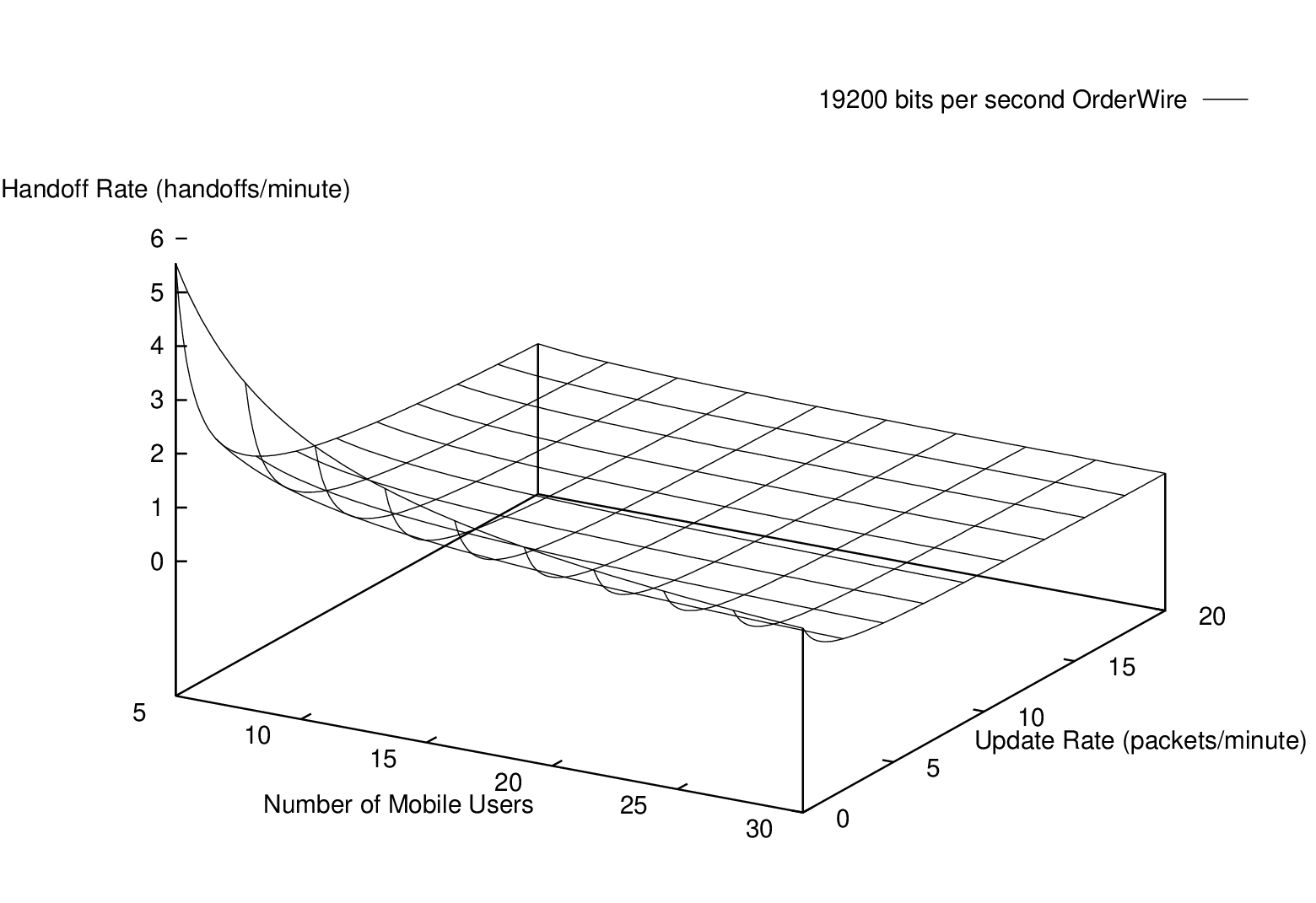,width=6.0in}}
        \caption{Orderwire Traffic Analysis {\bf with} Virtual Network Configuration.}
        \label{vncbw}
\end{figure*}

An estimate of the additional load on the orderwire packet radios
using VNC is shown in Figure \ref{vncbw}. It is assumed that virtual 
messages are 65 bits 
longer than real messages and there is one virtual message for each real
message. The figure shows the prototype 19,200 bps orderwire
link capacity as a function of the number of RNs, the position update
rate of each RN, and the hand-off rate. The capability of the orderwire 
to support these rates without
VNC is discussed later in detail and is shown in
Figure \ref{owcaphandreal}. Comparing Figures \ref{vncbw} and
\ref{owcaphandreal}, it is apparent that the VNC slightly more than
doubles the orderwire load. However enough capacity remains to support
users with a reasonable position update rate and handoff rate with 
this relatively low 19,200 bps orderwire bandwidth.

%% file: w5.tex
\subsection{Seamless Mobile ATM Routing}

\begin{figure*}[htbp]
        \centerline{\psfig{file=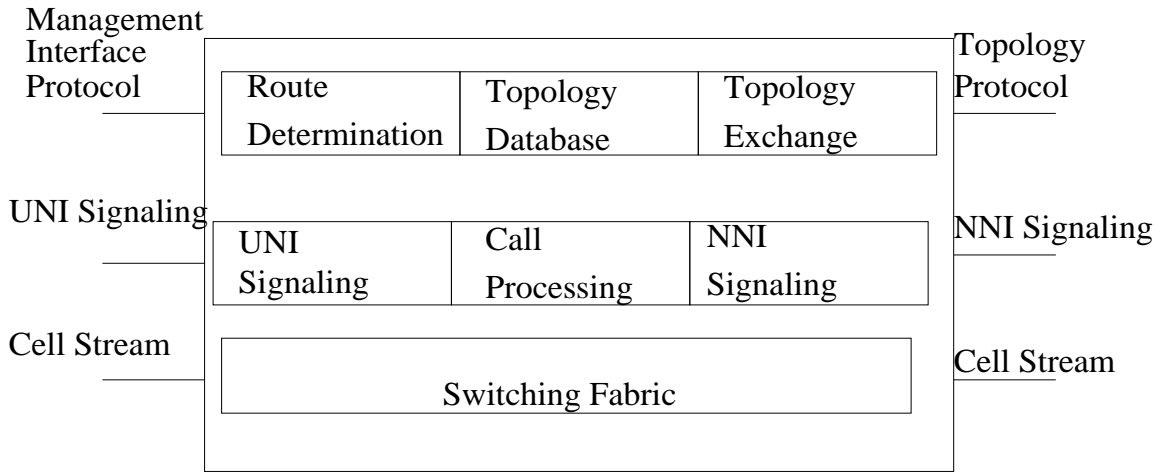,width=6.0in}}
        \caption{PNNI Architecture Reference Model.}
        \label{pnniov}
\end{figure*}

\begin{figure}[htbp]
        \centerline{\psfig{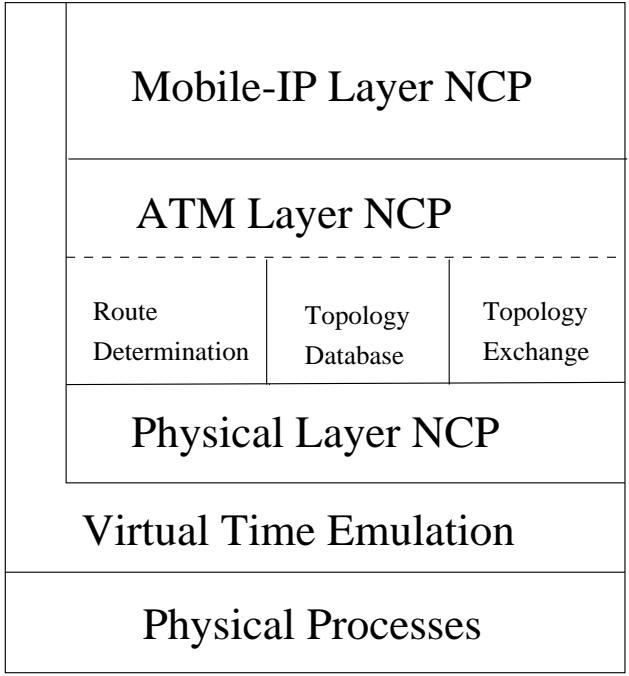}}
        \caption{Virtual Network Configuration Stack.}
        \label{vncstack}
\end{figure}

This section discusses an incorporation of Virtual Network Configuration 
(VNC) and the Network Control Protocol (NCP) as described in the previous 
sections into the Private Network-Network Interface (PNNI) \cite{PNNI} to 
facilitate seamless ATM hand-off. An attempt is made to minimize the changes 
to the evolving PNNI standard. Figure \ref{pnniov} shows a high level view 
of the PNNI Architecture. Terminology used in the PNNI Specification.
In this version of mobile PNNI, the standard PNNI route determination, 
topology database, and topology exchange would reside within the NCP.
The NCP stack with VNC is shown in Figure \ref{vncstack}.

The enabling mechanism is the fact that VNC will cause the NCP
to create a topology which will exist after a hand-off occurs at
a time prior to the hand-off.
This will cause PNNI to perform its standard action of updating
its topology information immediately before the hand-off occurs.
Note that this is localized within a single Peer Group (PG).

The second enabling mechanism is a change to the PNNI signaling protocol. 
In mobile PNNI, standard PNNI signaling is allowed to dynamically modify 
logical links when triggered by a topology change. This is similar to a
{\bf CALL ABORT} message except that the ensuing {\bf RELEASE} messages
will be contained within the scope of the Peer Group (PG). This new
message will be called a {\bf SCOPED CALL ABORT} message.

When the topology changes due to an end system hand-off, a check is 
made to determine which end system (RN) has changed logical nodes (LN). 
An attempt is made to establish the same incoming VCs at the new LN as were 
at the original LN and connections are established from the new LN to the 
original border LNs of the Peer Group. This allows the RN to continue 
transmitting with the same VCI as the hand-off occurs. The connections 
from the original LNs to the border LNs are released after the hand-off 
occurs. If the new LN is already using a VCI that was used at the original 
LN, the {\bf HANDOFF} packet will contain the replacement VCIs to be used by
the end system (RN). There are now two branches of a logical link tree 
established  with the border LN as the root. After the hand-off takes 
place the old branch is removed by the new {\bf SCOPED CALL ABORT} message.

Note that link changes are localized to a single Peer Group.
The fact that changes can be localized to a Peer Group greatly reduces 
the impact on the network and implies that the mobile network should have 
many levels in its PNNI hierarchy. In order to maintain cell order the 
new path within the Peer Group is chosen so as to be equal to or longer 
than the original path based on implementation dependent metrics.

\begin{figure*}[htbp]
        \centerline{\psfig{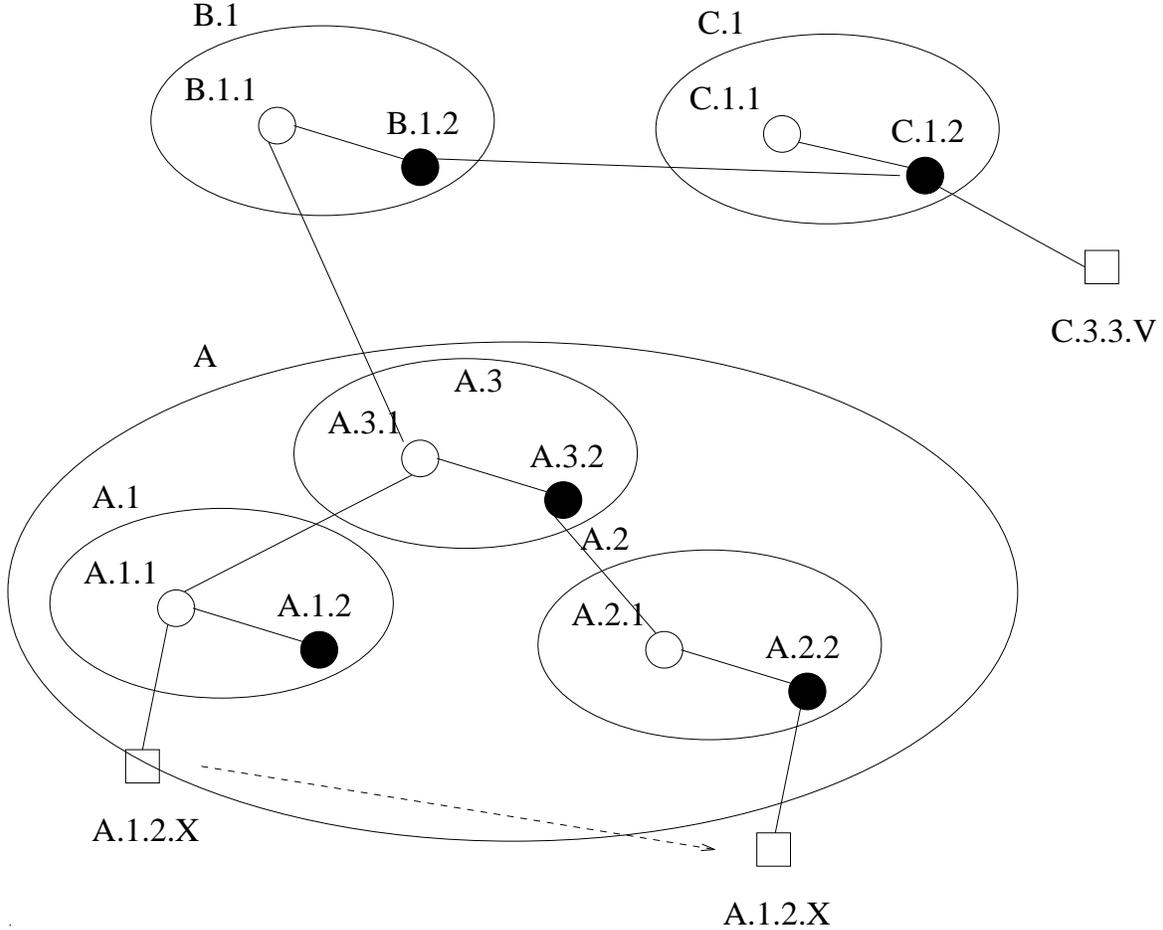}}
        \caption{Mobile PNNI Example.}
        \label{mobpnni}
\end{figure*}

Consider the network shown in Figure \ref{mobpnni}. 
Peer Groups are enclosed in circles and the blackened nodes
represent the lowest level Peer Group Leader for each Peer Group.
End system A.1.2.X is about to hand-off from A.1.1 to A.2.2. The smallest 
scope which encompasses the old and new LN is the LN A.

A.3.1 is the outgoing border node for LN A. A {\bf CALL SETUP} uses
normal PNNI operations to setup a logical link from A.3.1 to A.2.2.
After A.1.2.X hands off, a {\bf SCOPED CALL ABORT} message releases the 
logical link from A.3.1 to A.1.1.

%% file: s9.tex
\section{Development and Implementation}
\label{emulation}

The initial physical layer network control protocol design was done 
using Maisie \cite{bagrodia}, 
a C-based parallel programming language. It facilitates creation 
of entities which execute in parallel and the ability to easily 
send and receive messages between entities. 
A Maisie emulation of the entire network was developed which uses the actual
NCP code. This helped build confidence 
that the design of the Network Control Protocol was correct.

The network control protocol code was initially tested with only the two 
packet radios available. Since at least three packet radios are necessary 
for a complete RN-ES-RN orderwire connection, the next step involved 
emulating the packet radios via TCP/IP over Ethernet, and completing 
the development of the code. The packet radio emulation also allowed 
testing of various configurations that helped determine if the network control protocol
was scalable.

The physical layer of the Network Control Protocol is a single-unit
consumable resource system. There can be no
deadlock since there are no cycles.  All message  interactions take place 
with a master switch,
except for the initial {\bf MYCALL} packet broadcast.

The GPS system was also emulated to provide the appearance of
mobility so that hand-offs of a host from one ES to another
could be tested. The GPS emulation is also an important component of 
the Virtual Network Configuration Algorithm. The actual orderwire code is
used in these emulations.

%% file: s7.tex
\subsection{Timing Results}
\label{timres}

\begin{figure*}[htbp]
        \centerline{\psfig{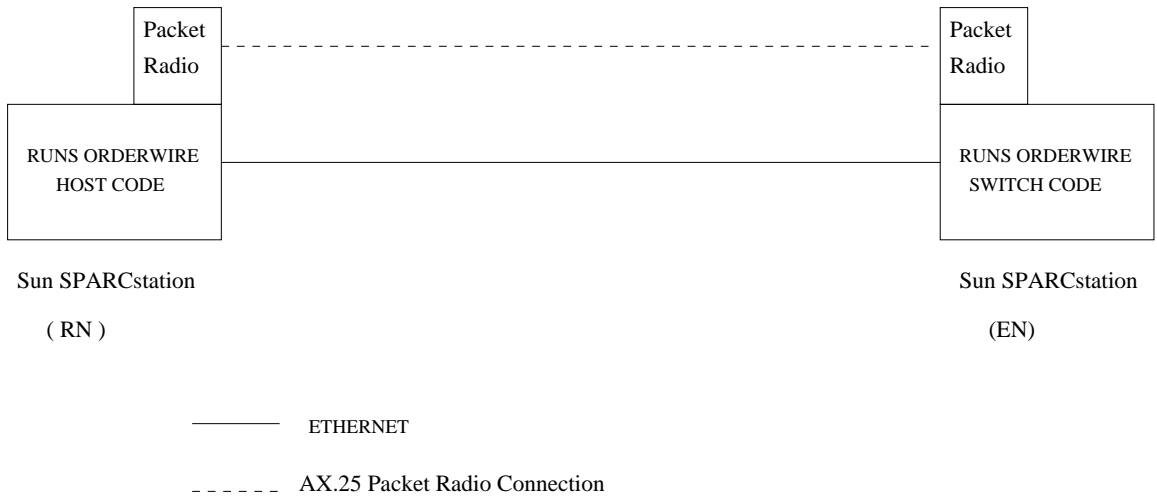}}
        \caption{Physical Setup for Packet Radio Timing.}
        \label{timepr}
\end{figure*}

This section summarizes the results of initial timing experiments that were 
undertaken to examine the performance of the orderwire system. The 
experiments involved determining the time required to transmit and process 
each of the packet types listed in Table \ref{prototab} using the real packet 
radios. These times represent the time to packetize, transmit, receive,
and depacketize each packet at the Network Control Protocol process.
Figure \ref{timepr} illustrates the physical configuration used for the 
experiments involving the real packet radios. The results are presented in 
Table \ref{timetab}. 
Most of the overhead occurs during the initial system configuration which
occurs only once as long as ESs remain stationary.
With regard to a handoff, the 473 millisecond time to transmit and process 
the handoff packet is on the same order of time as that required to compute 
the beam angles and steer the beams. The following sections provide an 
analysis and discuss the impact of scaling up the system on the configuration 
time.

\begin{table}
\centering
\begin{tabular}{||l|c||}                    \hline
{\bf Event}   & {\bf Time (ms)}          \\ \hline \hline
%MOBAGENT      & 469                 \\ \hline
%ARPSERVER     & 494                 \\ \hline
%AN-2 VC Setup & 300                 \\ \hline
%VC\_SETUP     & 476                 \\ \hline
USER\_POS     & 677                 \\ \hline
NEWSWITCH     & 439                 \\ \hline
HANDOFF       & 473                 \\ \hline
MYCALL        & 492                 \\ \hline
SWITCHPOS     & 679                 \\ \hline
TOPOLOGY      & 664                 \\ \hline
\end{tabular}
\caption{\label{timetab} Network Control Protocol Timing Results.}
\end{table}

%% file: s10.tex
\subsection{Bandwidth required for the Orderwire Network}
\label{owbw}

\begin{figure*}[htbp]
        \centerline{\psfig{file=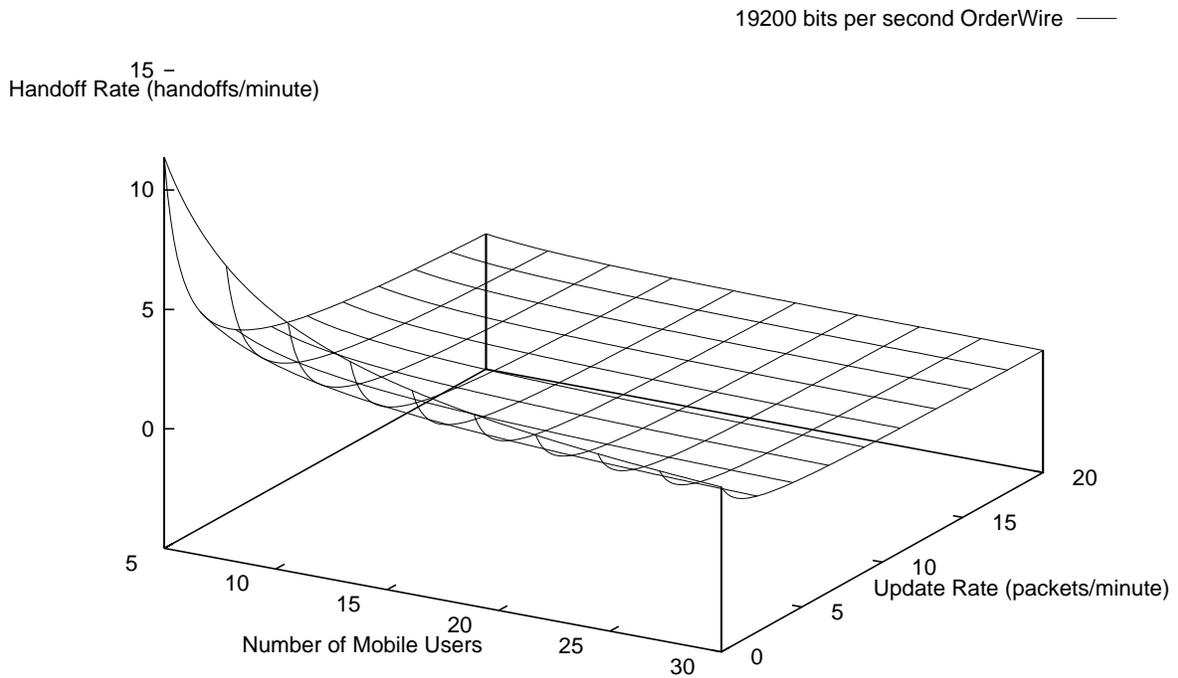,width=6.0in}}
        \caption{Orderwire Traffic Analysis {\bf without} Virtual Network Configuration.}
        \label{owcaphandreal}
\end{figure*}

 The traffic over the orderwire was analyzed to determine a relation between 
the maximum update rate and the number of RNs. The protocol used for contention
 resolution on the broadcast channel is the Aloha Protocol which is known to 
have a maximum efficiency close to 18\%. Given the bandwidth of the orderwire 
channel, size of an orderwire packet and this value for the efficiency, we 
compute and plot the value for the maximum update rate (in packets per minute) 
for a given number of RNs. The plot of Figure \ref{owcaphandreal} shows the 
variation in update rate for between 5 and 30 RNs. This study gives us 
an upper limit on the number of RNs that can be supported over the orderwire 
given a minimum required update rate and handoff rate.

%% file: w2.tex
\section{NCP Performance Analysis}
\label{NCPanal}

The analysis of the RDRN network configuration time using the protocols
proposed earlier, will be divided into three phases. 
Phase I is the ES-ES configuration, Phase II
is the RN configuration, and Phase III is handoff configuration.
The specific numerical values used in this section were obtained from
Table \ref{timetab}.

\subsection{Phase I}

In Phase I the ES nodes act in a distributed manner to determine which ES 
will become the master ES. The 
master ES collects position information, determines the optimum
ES interconnections, and distributes the results back to the
ES nodes. The ES nodes determine the master ES by broadcasting {\bf MYCALL}
packets and collecting {\bf MYCALL} packets until the
MYCALL Timer expires with a prespecified time, $T$. The {\bf MYCALL}
packets contain the callsign and boot-time
of each ES. The ES with oldest boot-time is designated as the master.
$T$ should be chosen as the smallest value
which allows enough time for all {\bf MYCALL} packets to be received.
This would be approximately $0.492 * (N-1)$ seconds, where $N$ is the 
total number of ES nodes. {\bf NEWSWITCH} 
packets take on the order of 0.439 seconds to transfer, and therefore, it 
will take $0.439 * (N-1)$ 
seconds to send these packets. 
The ES nodes will respond with {\bf SWITCHPOS} packets which will take
another $0.679 * (N-1)$ seconds. 
These events occur after each {\bf MYCALL} packet has been received,
and can occur before the MYCALL Timer has expired.

The next step in Phase I is to run the topology algorithm which is
based on a consistent labeling algorithm \cite{cla}. This algorithm
generates all fully connected topologies given ES node locations and
constraints on the antenna beams such that beams do not interfere with
one another. The information required by the topology module is the
GPS location of all ES nodes, transmit and receive beam widths, transmit
radius, the number of non-interfering frequency pairs, and an interference 
multiplier. An interference multiplier of 1.0 assumes adaptive power
control, in which case it is assumed that beam power control will be 
adjusted to exactly match the link distance. The interference multiplier
multiplied by a link's actual length will determine the range of
interference created by the link.
This takes on the order of $K_{top} \left[N^2+(L+1)^R \right]$ seconds
where $L$ is the number of available frequency pair
combinations with the addition of $1$ for no link. Assigning distinct
frequencies allows beams to overlap without interference.
$R$ is the number 
of constrained links and $K_{top}$ is a constant. The constraints are 
based on maximum beam length, beam widths, and number of frequencies
which can be supported. 

The final step in Phase I is to distribute the topology information
to all ES nodes in {\bf TOPOLOGY} packets. This takes approximately
$0.664 + (0.1 * (N-1))$ seconds.

The time for Phase I to complete as a function of N is shown in Equation \ref{timep1}.

\begin{eqnarray}
P1(N) & = & \max\left[T, 0.439 * (N-1) + 0.492 * (N-1)\right] + \nonumber \\
      &   & K_{top}\left[N^2+(L+1)^R\right] + \label{timep1} \\
      &   & 0.664 + (0.1 * (N-1)) \nonumber
\end{eqnarray}

\subsection{Phase II}

Phase II is the RN configuration phase. Let U be the number of RNs
associated with a given ES. The first step is for the ES to receive
{\bf USER\_POS} packets from each RN. This takes $0.677 * U$ seconds.

The next step is to determine the optimum direction of the beams in order
to form a connection with the RNs. This algorithm execution is a linear
function of the number of RNs,  which takes $K_{bf} * U$ seconds where 
$K_{bf}$ is a constant.  The algorithm is currently implemented in 
MatLab and takes approximately 7.5 seconds to obtain reasonable convergence 
of the beam direction to connect with four RNs.

The final step in Phase II is to generate a table of complex weights
for antenna beamforming and download this table to the hardware.
This is a function of the number of elements in the antenna
array, $K_{el}$, the number of beams, $B$, and the number of bits per
symbol, $M$. $K_{el}$ tables are created with $2^{M * B}$ entries per table.
This takes on the order of 2 seconds with 4 beams and 8 elements for
QPSK modulation on an OSF1 V4.0 386 DEC 3000/400 Alpha workstation.

The entire beamform and table generation module must be repeated for
every combination of transmitting RNs. A different table is used
depending on which RNs are currently transmitting data. 
The complete time for Phase II is shown in Equation \ref{p2}.

\begin{equation}
P2(U) = 0.677 * U + \sum_{r=1}^{U} {U \choose r} \left( K_{bf} * r + 
K_{el} * 2^{M * B} \right)
\label{p2}
\end{equation}

\subsection{Phase III}

Phase III, shown in Equation \ref{p3}, is the time required for the 
orderwire to perform a hand-off.
The current network control code determines RN to ES associations based
on distance. When the distance between an RN and an ES other than its 
currently associated ES becomes smaller than the distance between the
RN and its currently associated ES,
the current ES initiates a hand-off by sending a {\bf HANDOFF} packet.
This takes $0.473$ seconds. The RN will then initiate a point-to-point 
orderwire connection with the new ES. Finally, Phase II must be run again 
at the new ES, which is the reason for including the function $P2$.

\begin{equation}
P3_{RN} = 0.473 + P2(U+1)
\label{p3}
\end{equation}

%% file: w3.tex
\subsection{Orderwire Performance Emulation}

%\lgrindfile{/projects/rdrn/owcode/maisie_code/pr.tex}

The emulation of the orderwire systems satisfies several goals. It 
allows tests of configurations
that are beyond the scope of the prototype RDRN hardware. Specifically,
it verifies the correct operation of the RDRN Network Control Protocol 
in a wide variety of situations. 
The emulation helps to verify the correctness of the analytical 
results obtained above. 
As an additional benefit, much of the actual orderwire code was used by 
the Maisie \cite{bagrodia} emulation allowing further validation of that
code.
The Edge Switch (ES) and Remote Node (RN) are modeled as a collection of 
Maisie entities. This 
is an emulation rather than a simulation because the Maisie code is 
linked with the working orderwire code and also with the topology 
algorithm. 
There is an entity for each major component of the RDRN system
including the GPS receiver, packet radio, inter ES links, RN to ES links and
the Master, ES, and RN network configuration processors, as
well as other miscellaneous entities. 
The input parameters to the emulation are shown in Tables \ref{emin_move},
\ref{emin_time}, \ref{emin_beam}.

\begin{table*}[htbp]
\centering
\begin{tabular}{||l|l||}                                           \hline
{\bf Parameter }  & {\bf Definition}         \\ \hline \hline
NumRN		& Number of Remote Nodes  \\ \hline
NumES 		& Number of Edge Switches  \\ \hline
ESDist 		& Inter Edge Switch spacing (forms rectangular area) \\	\hline
T 		& ES/ES MYCALL configuration time \\ \hline
maxV 		& Maximum RN speed for uniform distribution \\ \hline
S 		& Time to wait between node initial startups \\ \hline
ESspd 		& Initial Edge Switch speed \\ \hline
ESdir 		& Initial Edge Switch direction \\ \hline
RNspd 		& Initial Remote Node speed \\ \hline
RNdir 		& Initial Remote Node direction \\ \hline
\end{tabular}
\caption{\label{emin_move} NCP Emulation Mobility Input Parameters.}
\end{table*}

\begin{table*}[htbp]
\centering
\begin{tabular}{||l|l||}                                           \hline
{\bf Parameter }  & {\bf Definition}         \\ \hline \hline
EndTime         & Emulation end time (tenths of seconds) \\ \hline
VCCallTime      & Inter High Speed Connection Setup Times \\ \hline
VCCallDuration  & High Speed Connection Life Times  \\ \hline
\end{tabular}
\caption{\label{emin_time} NCP Emulation Time Input Parameters.}
\end{table*}

\begin{table*}[htbp]
\centering
\begin{tabular}{||l|l||}                                           \hline
{\bf Parameter }  & {\bf Definition}         \\ \hline \hline
UseRealTopology & Connect to MatLab and run actual program \\ \hline
Rlink           & Maximum beam distance \\ \hline
Fmax            & Number of non-interfering frequency pairs \\ \hline
Imult           & Interference multiplier \\ \hline
Twidth          & Transmitting Beam width \\ \hline
Rwidth          & Receiving Beam width  \\ \hline
\end{tabular}
\caption{\label{emin_beam} NCP Emulation Beam Input Parameters.}
\end{table*}

The RN VC setup process for connections over the inter ES antenna beams
is assumed to be Poisson. This represents ATM VC usage over the
physical link. The RN will maintain a
constant speed and direction until a hand-off occurs, then a new
speed and direction are generated from a uniform distribution.
This simplifies the analytical computation. Note that NCP packet transfer 
times as measured in Table \ref{timetab} are used here.

\subsection{Orderwire Maisie Emulation Design}

\begin{figure*}[htbp]
        \centerline{\psfig{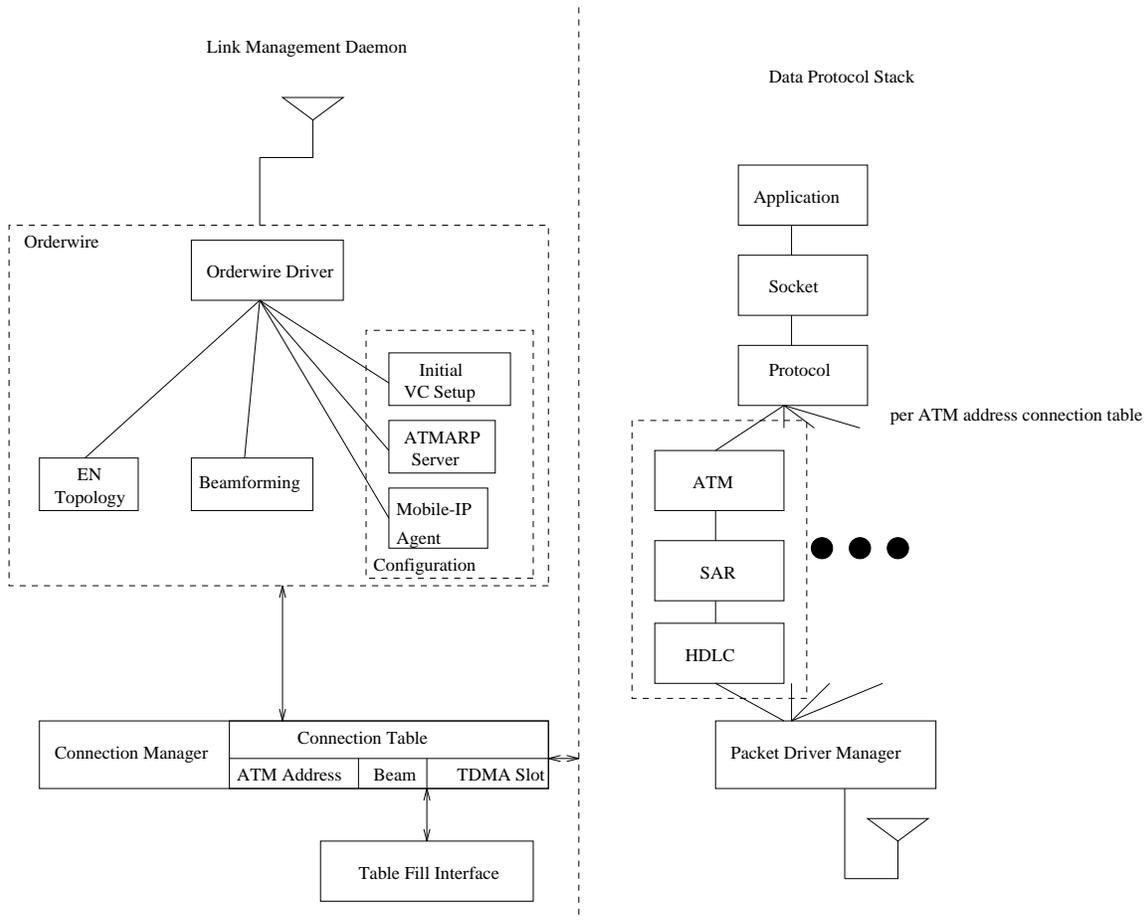}}
        \caption{Network Control System Architecture.}
        \label{ncsa}
\end{figure*}

The architecture for the RDRN link management and control is shown in
Figure \ref{ncsa}. The topology modules are used only on ES nodes capable of
becoming a master ES. The remaining modules are used on all ES nodes and RNs.
The beamform module determines an optimal steering
angle for the given number of beams which connects all RNs to be associated
with this ES. It computes an estimated signal to noise interference 
ratio (SIR) and generates a table of complex weights which, once loaded, 
will control the beam formation. Note that this table is not loaded until the table fill
trigger is activated. The connection table is used by the Adaptive HDLC and
ATM protocol stacks for configuration via the adaptation manager.

\begin{figure*}[htbp]
        \centerline{\psfig{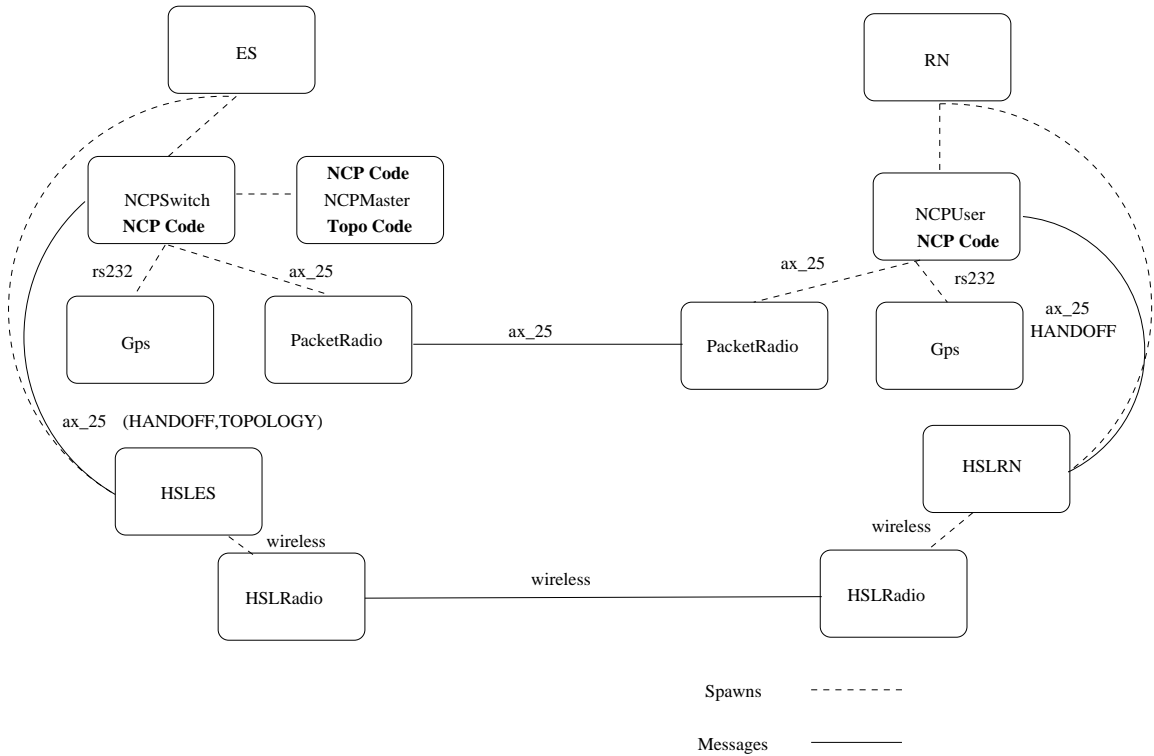}}
        \caption{Emulation Design.}
        \label{ncpem}
\end{figure*}

The emulation uses as much of the actual network control code as possible.
The packet radio driver, GPS driver, and Network Control Protocol state
machine are implemented in Maisie; tables, data structures, and decision
functions from working NCP code are used.
Figure \ref{ncpem} shows the structure of the Maisie entities. The entity
names are shown in the boxes and the message types are shown along the
lines. Direct communication between entities is represented as a solid line.
The dashed lines indicate from where entities are spawned.

%\onecolumn
%\small
%\lagrind[htbp]{code/pr.tex}{Packet Radio.}{pr}
%\normalsize
%\twocolumn

%\onecolumn
%\small
%\lagrind[htbp]{code/gps.tex}{GPS Entity.}{gps}
%\normalsize
%\twocolumn

The RN entity which performs ATM VC setup (HSLRN entity in Figure \ref{ncpem}) 
generates calls as a Poisson process which the ES node (HSLES entity in 
Figure \ref{ncpem}) will attempt to accept. If the EN moves out of range or the 
ES has no beam or slot available the setup will be aborted. As the RN moves,
the ES will hand off the connection to the proper ES based on closest distance 
between RN and ES. 

%% file: w4.tex
\section {Emulation Results}
\label{emres}

This section discusses the current results from the emulation. 
Some of these results revealed problems which are not immediately
apparent from the state diagrams in Figures \ref{pap1}, \ref{pap3},
\ref{pap2}. 
The emulation produces Network Control Protocol Finite State Machine 
(NCP FSM) output which shows the transitions based on the state diagrams 
in Figures \ref{pap1}, \ref{pap3}, \ref{pap2}. The FSM output provides
an easy comparison with diagrams to insure correct operation of the protocol.

\subsection{Effect of Scale on NCP}

The emulation was run to determine the effect on the NCP as the number
of ES and RN nodes increased. The dominate component of the configuration
time is the topology calculation run by the ES which is designated as the
master. Topology calculation involves searching through the problem space of
constraints on the directional beams for all feasible topologies and 
choosing an optimal topology from that set as described in \cite{cla}. 
The units on all values should be consistent with the GPS coordinate units, 
and all angles are assumed to be degrees.
The beam constraint values are Maximum link distance 1000.0, Maximum
Frequencies 3, Interference Multiplier 1.0, Transmit Beam Width 10.0,
Receive Beam Width 10.0. 

\begin{comment}
\begin{figure}[htbp]
        \centerline{\psfig{file=figures/top.ps,width=3.25in}}
        \caption{Topology Calculation Execution Time.}
        \label{toptime}
\end{figure}
\end{comment}

The topology calculation is performed
in MatLab and uses the MatLab provided external C interface. Passing
information through this interface is clearly slow, therefore
these results do not represent the exact execution times of the prototype
system. However, they do provide a worse case test for the protocol.

A possible speedup may arise through the use of Virtual Network 
Configuration, which will provide a mechanism for predicting values in 
advance and also allows processing to be distributed.
Another improvement which may be considered is to implement a 
hierarchical configuration. The network is partitioned into
a small number of clusters of nodes in such a way that nodes in each 
group are as close 
together as possible. The topology code is run as though these were
individual nodes located at the center of each group. This inter-group
connection will be added as constraints to the topology
computation for the intra-group connections. In this way the topology
program only needs to calculate small numbers of nodes which it does
relatively quickly.

\begin{comment}
\begin{figure}[htbp]
        \centerline{\psfig{file=figures/config.ps,width=3.25in}}
        \caption{Configuration Time.}
        \label{configtime}
\end{figure}
\end{comment}
 
\subsection{MYCALL Timer}

\begin{figure}[htbp]
        \centerline{\psfig{file=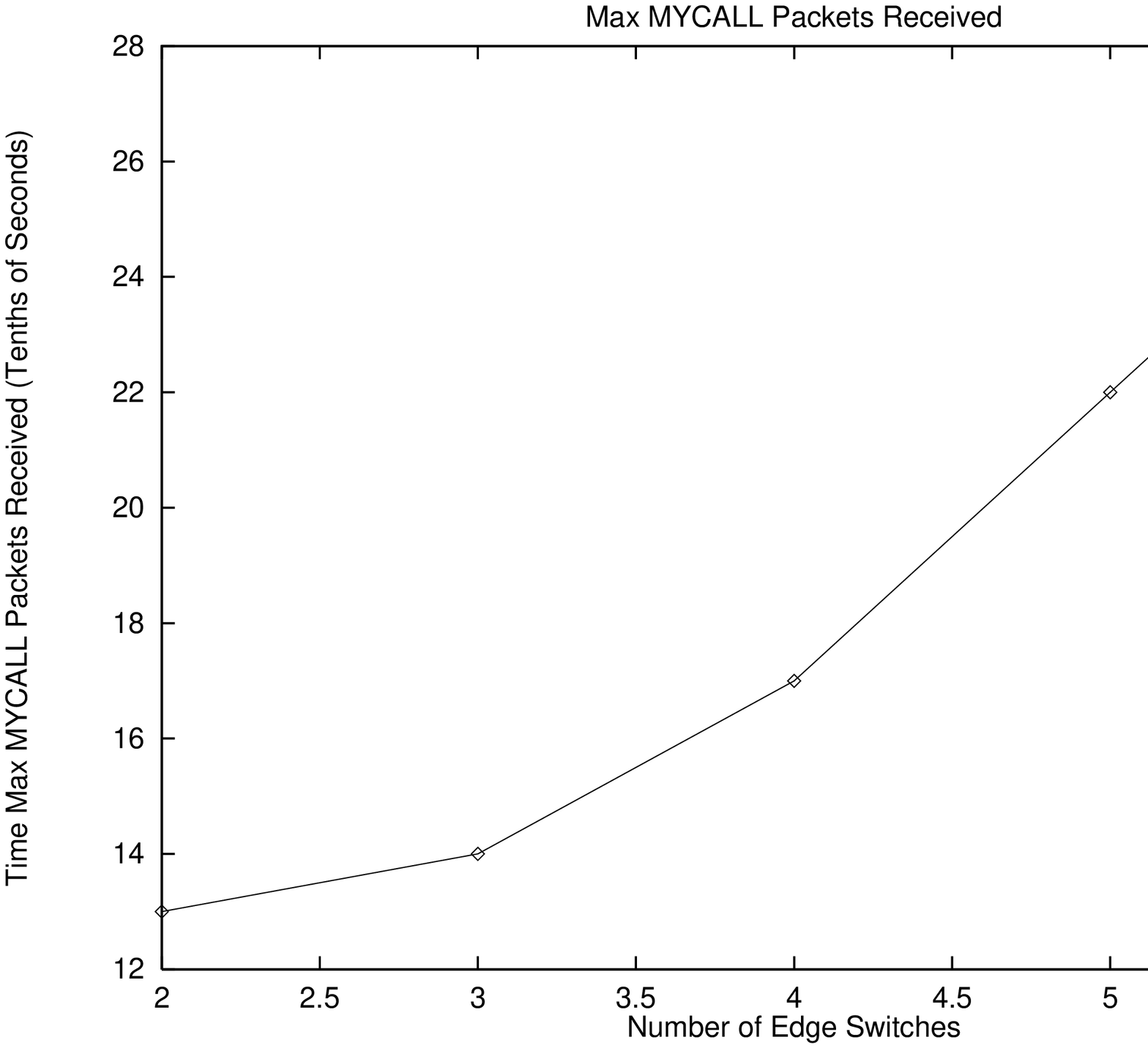,width=3.25in}}
        \caption{MYCALL Packets Received.}
        \label{mycalltime}
\end{figure}

The MYCALL Timer, set to a value of $T$ in the analysis section, controls
how long the system will wait to discover new ES nodes before completing the
configuration. If this value is set too low, new {\bf MYCALL} packets will
arrive after the topology calculation has begun, causing the system to 
needlessly reconfigure.
If the MYCALL Timer value is too long, time will be wasted, which will
have a large impact on a mobile ES system. Table \ref{mycallsim} shows the
input parameters and Figure \ref{mycalltime} shows
the time required for all {\bf MYCALL} packets to be received as a
function of the number of ES nodes. These times are the 
optimal value of the MYCALL Timer as a function of the number of
ES nodes because the these times are exactly the amount of time 
required for all ESs to respond.
In order to prevent the possibility of an infinite loop of reconfigurations
from occurring, an exponential back-off on
the length of the {\bf MYCALL} Timer value is introduced. 
As {\bf MYCALL} packets arrive after
T has expired, the next configuration occurs with an increased value
of T. 

\begin{table*}[htbp]                           
\centering                                                      
\begin{tabular}{||l|l||}                              \hline
{\bf Parameter }     & {\bf Value}                 \\ \hline 
Number of RNs        & $0$                         \\ \hline 
Number of ESs        & $2$ thru $6$                \\ \hline
Inter-ES Distance    & $20$ m (65.62 ft)           \\ \hline 
T                    & $20$ s                      \\ \hline
Maximum Velocity     & $5$ m/s (11.16 mi/hr)       \\ \hline
Initial ES Speed     & $0$ m/s                     \\ \hline
Initial ES Direction & $0^o$                       \\ \hline
Initial RN Speed     & N/A                         \\ \hline 
Initial RN Direction & N/A                         \\ \hline
Inter-VC Setup Time  & $1200$ s                    \\ \hline
VC Call Duration     & $600$ s                     \\ \hline
\end{tabular}
\caption{\label{mycallsim} MYCALL Timer Simulation Parameters.}
\end{table*}

\subsection{Link Usage Probability}

\begin{figure}[htbp]
        \centerline{\psfig{file=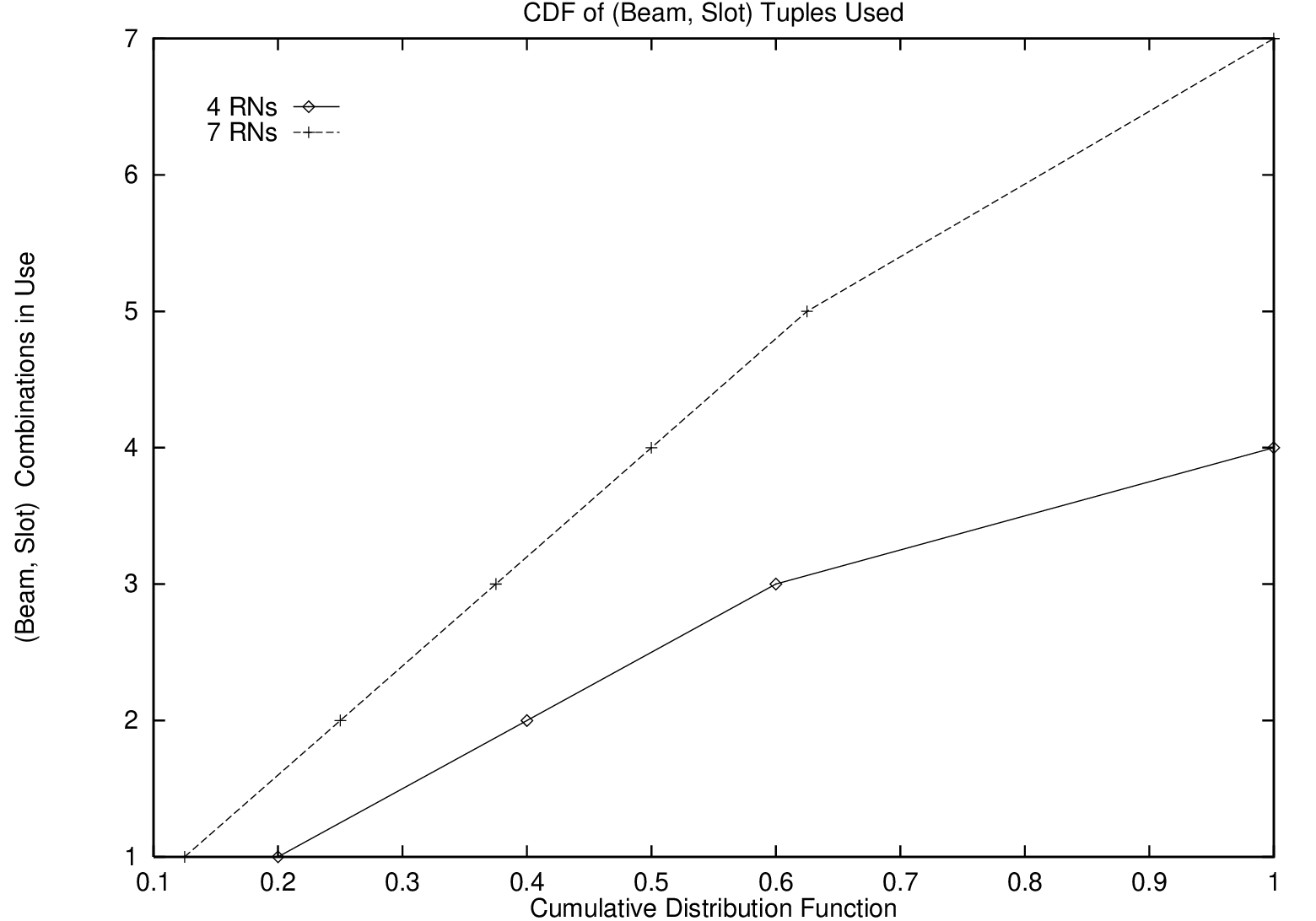,width=3.25in}}
        \caption{$(Beam, Slot)$ Usage.}
        \label{chanused}
\end{figure}

Multiple RNs may share a single beam using Time Division Multiplexing
(TDMA) within a beam. The time slices are divided into slots, thus
a $(beam, slot)$ tuple defines a physical link. The emulation was run
to determine the probability distribution of links used as a function
of the number of RNs. The parameters used in the emulation are shown in 
Table \ref{chanusedsim} the results of which indicate the number of links 
and thus the number of distinct $(beam, slot)$ tuples required. Figure 
\ref{chanused} shows the link usage cumulative distribution function
for 4 and 7 RNs. 

\begin{table*}[htbp]                           
\centering                                                      
\begin{tabular}{||l|l||}                              \hline
{\bf Parameter }     & {\bf Value}          \\ \hline \hline 
Number of RNs        & $4$ and $7$          \\ \hline
Number of ESs        & $2$                  \\ \hline
Inter-ES Distance    & $20$ m (65.62 ft)    \\ \hline
T                    & $20$ s               \\ \hline
Maximum Velocity     & $5$ m/s (11.16 mi/hr)\\ \hline
Initial ES Speed     & $0$ m/s              \\ \hline
Initial ES Direction & $0^o$                \\ \hline
Initial RN Speed     & $5$ m/s (11.16 mi/hr)\\ \hline 
Initial RN Direction & $0^o$                \\ \hline
Inter-VC Setup Time  & $1200$ s             \\ \hline
VC Call Duration     & $600$ s              \\ \hline
\end{tabular}
\caption{\label{chanusedsim} Link Usage Simulation Parameters.}
\end{table*}

\subsection{ES Mobility}

\begin{figure}[htbp]
        \centerline{\psfig{file=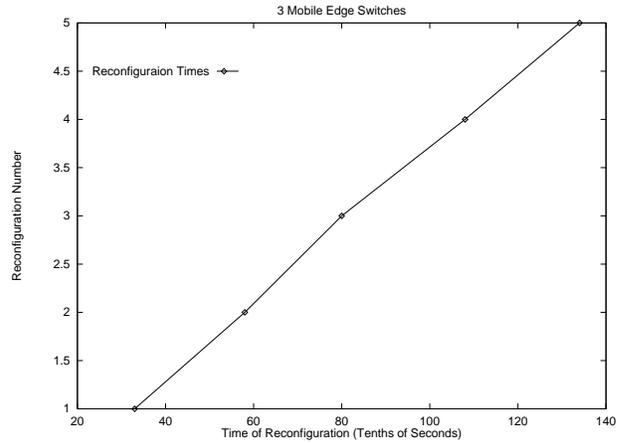,width=3.25in}}
        \caption{Mobile ES Configuration Time.}
        \label{mobestime}
\end{figure}

ES mobility is a more difficult problem and will be examined in more 
detail as the research proceeds. 
The parameters used in an emulation with mobile ES nodes are shown in Table
\ref{mobessim}.
As mentioned in the section on the MYCALL Timer, if a {\bf MYCALL} packet
arrives after this timer has expired, a reconfiguration occurs. This could
happen due to a new ES powering up or an ES which has changed position.
Figure \ref{mobestime} shows the times at which reconfigurations occurred
in a situation in which ES nodes were mobile.
Based on the state transitions generated from the emulation it is
apparent that the system is in a constant state of reconfiguration; no 
reconfiguration has time to complete before a new one begins. 
As ES nodes move, the NCP must notify RNs associated with an ES with the 
new position of the ES as well as reconfigure the ES nodes.
To solve this problem, a tolerance, which may be associated with the 
link quality, 
will be introduced which indicates how far nodes can move
within in a beam before the beam angle must be recalculated,
which will allow more time between reconfigurations.
It is expected that this tolerance in addition to Virtual Network 
Configuration will provide a solution to this problem.

\begin{table*}[htbp]                           
\centering                                                      
\begin{tabular}{||l|l||}                               \hline
{\bf Parameter }       & {\bf Value}                 \\ \hline \hline
Number of RNs          & $0$                         \\ \hline 
Number of ESs          & $3$                         \\ \hline 
Inter-ES Distance      & $20$ m (65.62 ft)           \\ \hline
T                      & $20$ s                      \\ \hline
Maximum Velocity       & $5$ m/s (11.16 mi/hr)       \\ \hline
Initial ES Speed       & $1$ m/s (2.23 mi/hr)        \\ \hline
Initial ES Direction   & $0^o$                       \\ \hline
Initial RN Speed       & N/A                         \\ \hline
Initial RN Direction   & N/A                         \\ \hline
Inter-VC Setup Time    & $1200$ s                    \\ \hline
VC Call Duration       & $600$ s                     \\ \hline
\end{tabular}
\caption{\label{mobessim} Mobile ES Simulation Parameters.}
\end{table*}

\subsection{Effect of Communication Failures}

The emulation was run with a given probability of failure on each
packet type of the Network Control Protocol. The following results
are based on the output of the finite state machine (FSM) transition 
output of the emulation
and an explanation is given for each case.

A dropped {\bf MYCALL}
packet has no effect as long as at least one of the {\bf MYCALL}
packets from each ES is received at the master ES. This is the
only use of the AX.25 broadcast mode in the ES configuration. The
broadcast AX.25 mode is a one time, best effort delivery;
therefore, {\bf MYCALL} packets are repeatedly broadcast at the
NCP layer.

The Maisie emulation demonstrated that
a dropped {\bf NEWSWITCH} packet caused the
protocol to fail. This is because the master ES will wait until
it receives all {\bf SWITCHPOS} packets from all ES nodes for 
which it had received {\bf MYCALL} packets. The {\bf NEWSWITCH}
packet is sent over the AX.25 in connection-oriented mode, e.g. 
a mode in which corrupted frames are retransmitted;
the probability of loosing a packet in this mode is 
very low.
A dropped {\bf SWITCHPOS} packet has the same effect as a dropped
{\bf NEWSWITCH} packet.
In order to avoid this situation, the NCP will re-send the {\bf NEWSWITCH}
if no response is received.

Finally, the Maisie emulation showed that
a lost {\bf TOPOLOGY} packet results in a partitioned network.
The ES which fails to receive the {\bf TOPOLOGY} packet is not joined with
the remaining ES nodes; however, this ES node continued to receive and 
process {\bf USER\_POS} packets from all RNs. It therefore attempts to 
form an initial connection with all RNs.
The solution for this condition is not to allow RN associations with
an ES node until the {\bf TOPOLOGY} packet is received. Because
{\bf MYCALL} packets are transmitted via broadcast AX.25, each ES node 
can simply count the number of {\bf MYCALL} packets and estimate the time for 
the master ES node to calculate the topology using the number of {\bf MYCALL}
packets as an estimate for the size of the network. If no {\bf TOPOLOGY} 
packet is received within this time period, the ES node retransmits its 
{\bf SWITCHPOS} packet to the master ES node in order to get a
{\bf TOPOLOGY} packet as a reply.

%% file: s6.tex
\section{Summary}
\label{conclusion}

This paper described the design of a control and management network 
for a mobile wireless ATM network. The orderwire system consists of a 
packet radio network which overlays the mobile wireless ATM network and 
receives GPS information. This information is used to control the 
beamforming antenna subsystem which provides for spatial reuse.
This paper also proposed the design of the VNC algorithm which is a 
novel concept for predictive configuration. A mobile ATM 
PNNI based on VNC was also discussed. 
As a prelude to the system implementation, results
of a Maisie simulation of the orderwire system were presented.
Finally, the Network Control Protocol was tested, initial problems
corrected, and initial performance results were obtained and
presented in this paper.